\documentclass[conference,10pt]{IEEEtran}
\usepackage{cite}
\usepackage{graphicx}
\usepackage{psfrag}
\usepackage{url}
\usepackage{amsmath}
\usepackage{array}
\usepackage{amssymb}
\usepackage{mathtools}
\usepackage{amsfonts}
\usepackage{graphicx}
\usepackage{epstopdf}
\usepackage{algorithm}
\usepackage{algorithmic}

\usepackage{setspace}
\usepackage{amscd}
\usepackage{mathrsfs}
\usepackage{epsfig}
\usepackage{color}
\usepackage{textcomp}
\usepackage{tikz}
\usetikzlibrary{shapes,snakes,plotmarks}

\usepackage{caption}
\usepackage{subcaption}

\newcommand{\boundellipse}[3]{(#1) ellipse (#2 and #3)}

\IEEEoverridecommandlockouts

\DeclareMathOperator{\diag}{diag}
\DeclareMathOperator{\trace}{Tr}
\DeclareMathOperator*{\argmax}{argmax}

\makeatletter
\newcommand\fs@spaceruled{\def\@fs@cfont{\bfseries}\let\@fs@capt\floatc@ruled
  \def\@fs@pre{\vspace{0.5\baselineskip}\hrule height.7pt depth0pt \kern2pt}%
  \def\@fs@post{\kern2pt\hrule\relax}%
  \def\@fs@mid{\kern2pt\hrule\kern2pt}%
  \let\@fs@iftopcapt\iftrue}
\makeatother

\title{An Algorithm for Grant-Free Random Access in Cell-Free Massive MIMO}

\begin{document}
\author{\IEEEauthorblockN{Unnikrishnan Kunnath Ganesan, Emil Bj\"ornson and Erik G. Larsson}
		\IEEEauthorblockA{Department of Electrical Engineering (ISY)\\
		Link\"oping University, Link\"oping, SE-581 83, Sweden.\\
		Emails: \{unnikrishnan.kunnath.ganesan, emil.bjornson, erik.g.larsson\}@liu.se}
\thanks{This work is supported in part by ELLIIT and in part by  Swedish Research Council (VR).}
}
\maketitle
\thispagestyle{empty}

%%%%%%%%
\begin{abstract}
Massive access is one of the main use cases of beyond 5G (B5G) wireless networks and massive MIMO is a key technology for supporting it. Prior works studied massive access in the co-located massive MIMO framework. In this paper, we investigate the activity detection in grant-free random access for massive machine type communications (mMTC) in cell-free massive MIMO network. Each active device transmits a pre-assigned non-orthogonal pilot sequence to the APs and the APs send the received signals to a central processing unit (CPU) for joint activity detection. We formulate the maximum likelihood device activity detection problem and provide an algorithm based on coordinate descent method having affordable complexity. We show that the cell-free massive MIMO network can support low-powered mMTC devices and provide a broad coverage.
\end{abstract}

\begin{IEEEkeywords} 
Activity Detection, Grant-Free Random Access, Cell-Free massive MIMO, massive machine-type communications (mMTC), Internet-of-Things (IoT).
\end{IEEEkeywords}

\section{Introduction}	
Massive machine type communications (mMTC) \cite{dutkiewicz2017massive} is one of the main requirements of future beyond 5G (B5G) wireless networks \cite{chen2020massive} and is an enabler for massive connectivity in Internet-of-Things (IoT). One of the main challenges of mMTC is that the network should be able to support a large number of devices over the same time and frequency resources while keeping battery lives of the devices as long as possible. Massive MIMO is a promising 5G technology to support massive access  \cite{marzetta2010noncooperative,bana2019massive}. 

Grant-based massive access is studied in \cite{pratas2012code, sorensen2014massive, bjornson2016random}. In the grant-based approach, each active device randomly picks a pilot or preamble sequence from a pool of orthogonal sequences, and uses the selected sequence to inform the base station that it has data to transmit. The base station needs to resolve collisions when they occur and a grant of resources will be provided to selected devices based on collision resolution. Due to the limited coherence interval, the set of orthogonal preamble sequences is finite. Grant-based protocols permit simple signal processing at the base station. A key feature of mMTC is that the traffic is sporadic with a very small fraction of potential devices being active and with very small payloads.  Thus, in the massive connectivity scenario, the probability of multiple active devices selecting the same sequence is quite high. Thus the grant-based protocols suffer from access failure due to collisions and hence increases the average latency. Also, due to collision resolution, the signaling overhead is quite large compared to the short payload each device has to send in mMTC applications. Thus, it is inefficient to use conventional grant-based access methods for mMTC. 

Various grant-free protocols have been proposed for the active devices to access the cellular wireless network without a grant. At the expense of sophisticated signal processing at the base station, the grant-free approach reduces the access latency and signaling overhead compared to grant-based approaches. In the grant-free approach, each device is assigned a unique preamble sequence and the active devices access the network using this preamble and the base station jointly decodes the devices which are active from the received signals. Due to the massive number of devices and limited coherence interval, preamble sequences are non-orthogonal and thus the received signal at the base station can suffer from severe co-channel interference. Thus activity detection is a challenging problem in the grant-free massive access scenario. Due to the sparse nature of the device activity pattern, the activity detection problem can be formulated as a compressive sensing (CS) problem and algorithms like approximate message passing (AMP) can be utilized \cite{chen2019multi,liu2018massive,liu2018sparse,senel2018grant}. However, performance of the CS based algorithms degrade severely when the number of active devices is larger than the coherence interval or the preamble sequence length. A covariance-based approach is proposed in \cite{haghighatshoar2018new} for device activity detection which performs better than CS based AMP schemes and an asymptotic performance analysis is given in \cite{chen2019covariance}. To have high success rate, multiple preamble based grant-free random access is studied in \cite{jiang2019multiple}. Activity detection in unsourced random access where all devices use the same codebook is studied in \cite{fengler2019grant}.

To provide high per-user data rates in B5G networks, one primary approach is densification of the network infrastructure by increasing the number of antennas per cell and deploying smaller cells. However, inter-cell interference is a limiting factor as we densify the network. Thus to overcome inter-cell interference, cell-free massive MIMO is a promising approach \cite{ngo2017cell}. Prior works study mMTC in the co-located massive MIMO architecture. In this paper, we investigate grant-free random access in cell-free massive MIMO networks. mMTC devices transmit signals with low power in order to keep the battery life as long as possible. Due to path loss and shadowing in wireless networks the transmitted signal attenuates heavily and the SNR at the base station reduces significantly. Cell-free MIMO networks can provide better coverage due to shorter propagation distances and is more robust to shadow fading effects compared to co-located MIMO at the expense of increased cost of fronthaul requirements. The authors in \cite{xu2015active} propose a Bayesian CS based algorithm exploiting the channel information and chunk sparsity structure for device activity detection in cloud radio access networks but it has huge computational complexity due to matrix inversions of high dimension. 

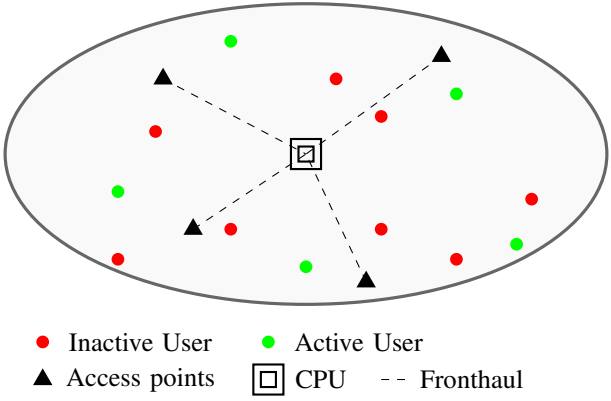
\begin{figure}[t]
	\centering
	\begin{tikzpicture}
	\filldraw[color=black!60, fill=gray!5, very thick](0,0) \boundellipse{0,0}{4}{2};
	\filldraw[red, thick] (-1,-1) circle (2pt);
	\filldraw[red, thick] (1,0.5) circle (2pt);
	\filldraw[red, thick] (-2,0.3) circle (2pt);
	\filldraw[red, thick] (0.4,1) circle (2pt);
	\filldraw[red, thick] (1,-1) circle (2pt);
	\filldraw[red, thick] (2,-1.4) circle (2pt);
	\filldraw[red, thick] (-2.5,-1.4) circle (2pt);
	\filldraw[red, thick] (3,-0.6) circle (2pt);
	
	\filldraw[green, thick] (-2.5,-0.5) circle (2pt);	
	\filldraw[green, thick] (2,0.8) circle (2pt);	
	\filldraw[green, thick] (0,-1.5) circle (2pt);	
	\filldraw[green, thick] (-1,1.5) circle (2pt);	
	\filldraw[green, thick] (2.8,-1.2) circle (2pt);	
	
	\node[fill=black,regular polygon, regular polygon sides=3,inner sep=1.5pt] at (-1.9,1) {};
	\draw[-, dashed] (-1.9,1) -- (0,0);
	\node[fill=black,regular polygon, regular polygon sides=3,inner sep=1.5pt] at (-1.5,-1) {};	
	\draw[-, dashed] (-1.5,-1) -- (0,0);
	\node[fill=black,regular polygon, regular polygon sides=3,inner sep=1.5pt] at (1.8,1.3) {};
	\draw[-, dashed] (1.8,1.3) -- (0,0);
	\node[fill=black,regular polygon, regular polygon sides=3,inner sep=1.5pt] at (0.8,-1.7) {};
	\draw[-, dashed] (0.8,-1.7) -- (0,0);	
	
	\draw[black, thick] (-0.2,-0.2) rectangle (0.2,0.2);
	\draw[black, thick] (-0.1,-0.1) rectangle (0.1,0.1);
	
	\filldraw[red, thick] (-3.5,-2.5) circle (2pt);
	\node at (-2.2,-2.5) {Inactive User};
	\filldraw[green, thick] (-0.5,-2.5) circle (2pt);
	\node at (0.7,-2.5) {Active User};
	\node[fill=black,regular polygon, regular polygon sides=3,inner sep=1.5pt] at (-3.5,-3) {};	
	\node at (-2.2,-3) {Access points};
	\draw[black, thick] (-0.7,-3.2) rectangle (-0.3,-2.8);
	\draw[black, thick] (-0.6,-3.1) rectangle (-0.4,-2.9);
	\node at (0.2,-3) {CPU};
	
	\draw[-, dashed] (1,-3) -- (1.4,-3);
	\node at (2.2,-3) {Fronthaul};
	\end{tikzpicture}
	\caption{Cell-Free Network Model for massive communication}
	\label{fig:network}
	\vspace{-15pt}
\end{figure}

In this paper, we formulate the maximum likelihood activity detection problem and provide an approximate solution for the estimator that has affordable complexity. An algorithm based on coordinate descent method is provided for activity detection in cell-free massive MIMO networks. We show that the cell-free massive MIMO network can support low-powered mMTC devices and is robust against shadow fading effects and hence can provide a broad coverage.

\textbf{\textit{Notations:}} Bold, lowercase letters are used to denote vectors and bold, uppercase letters are used to denote matrices. $\mathbb{C}$ denote the set of complex numbers. For a matrix $\mathbf{A}$, $\mathbf{A}^T$ and $\mathbf{A}^{H}$ denotes transpose and Hermitian transpose of the matrix $\mathbf{A}$ respectively. $\mathcal{CN}(0,\sigma^2)$ denotes a circularly symmetric complex Gaussian random variable with zero mean and variance equal to $\sigma^2$. $\mathbf{I}_N$ and $\mathbf{O}_N$ represents the $N \times N$ identity matrix and null matrix respectively. $|\mathcal{X}|$ denotes the cardinality of set $\mathcal{X}$.

\section{Signal Model And Problem Formulation}
\label{sec:SignalModel}

Consider a cell-free massive MIMO wireless network as illustrated in Fig. \ref{fig:network} with $M$ uniformly distributed APs each equipped with $N$ antennas serving $K$ uniformly distributed single antenna users. All the $M$ access points are assumed to be connected to a central processing unit (CPU) through a lossless infinite capacity fronthaul. Due to sporadic traffic in the massive access scenario of mMTC, only a small fraction of users are active at any time instant. In this paper, we assume that all users are synchronized during transmission and each user transmits independently with an activation probability $\epsilon$. Let $a_k\in\{0,1\}$  where $a_k =1$ denotes that the $k^{th}$ device is active and $a_k = 0$ that it is inactive and $\text{Pr}(a_k=1)=\epsilon$ and $\text{Pr}(a_k=0)=1-\epsilon$. The set of active users is denoted by $\mathcal{A}$ i.e., $\mathcal{A} = \{k : a_k=1\}$.  The channel between the $n^{th}$ antenna in the $m^{th}$ AP to device $k$ is given by 
\begin{equation}
g_{mnk} = \beta_{mk}^{\frac{1}{2}}h_{mnk}
\end{equation}
where $\beta_{mk}$ is the large-scale fading coefficient between the $m^{th}$ AP and user $k$ and $h_{mnk} \sim \mathcal{CN}(0,1)$ is the small-scale fading coefficient. We assume that the large-scale fading coefficient parameters $\{\beta_{mk}\}$ are available at all the transceivers. Throughout this paper, we consider a block fading scenario where the channel remains constant during the coherence interval \cite[Ch.2]{marzetta2016fundamentals} and all the channels are independent. Due to large number of users, typically $K\gg \tau_c$, where $\tau_c$ is the number of channel uses per coherence interval, assigning orthogonal pilots to each user is not feasible. Instead we assign non-orthogonal unique signature sequences $\mathbf{s}_k\in \mathbb{C}^{L\times1}$ to each user $k$ where $L\leq\tau_c$. The signature sequence are generated in an independent and identically distributed manner, i.e., $\mathbf{s}_{k} \sim \mathcal{CN}(0,\mathbf{I}_L) , \forall k $. We assume that all the devices are synchronized during transmission which means in an orthogonal frequency division multiplexing system, the time delays from different devices are well within the cyclic prefix. 

The $\mathbb{C}^{L\times 1}$ signal received at the $n^{th}$ antenna of $m^{th}$ AP is given by 
\vspace{-5pt}
\begin{equation}
\begin{aligned}
\mathbf{y}_{mn} & = \sum_{k=1}^{K}a_k\rho_k^{\frac{1}{2}} g_{mnk} \mathbf{s}_k + \mathbf{w}_{mn} \\
& = \mathbf{SD_aD}_{\boldsymbol{\rho}}^{\frac{1}{2}}\mathbf{g}_{mn}  + \mathbf{w}_{mn}
\end{aligned}
\end{equation}
where $\mathbf{S}=[\mathbf{s}_1 \ \mathbf{s}_2 \ \dots \ \mathbf{s}_K] \in \mathbb{C}^{L\times K}$ is the collection of all signature sequences, $\rho_k$ is the power transmitted by the user $k$, $ \mathbf{D_a} = \diag(a_1,a_2,\dots,a_K)$, $ \mathbf{D}_{\boldsymbol{\rho}} = \diag(\rho_1,\rho_2,\dots,\rho_K)$, $\mathbf{g}_{mn} = [g_{mn1} \ g_{mn2} \ \dots \ g_{mnK}]^T \in \mathbb{C}^{K\times1}$ is the channel vector from all $K$ users to the $n^{th}$ antenna of the $m^{th}$ AP and $\mathbf{w}_{mn}\sim \mathcal{CN}(\mathbf{0},\sigma^2\mathbf{I}_L)$ is the additive white Gaussian noise vector.

Thus, the $\mathbb{C}^{L\times N}$ signal received at the $m^{th}$ AP can be expressed as 
\vspace{-5pt}
\begin{equation}
\label{Ym_BS}
\mathbf{Y}_m = \mathbf{SD_aD}_{\boldsymbol{\rho}}^{\frac{1}{2}}\mathbf{G}_{m}  + \mathbf{W}_{m}
\end{equation}
where $\mathbf{G}_m = [\mathbf{g}_{m1} \ \mathbf{g}_{m2} \ \dots \ \mathbf{g}_{mN}]\in \mathbb{C}^{K\times N}$ is the channel matrix between the $K$ users and the $m^{th}$ AP and $\mathbf{W}_{m} = [\mathbf{w}_{m1} \ \mathbf{w}_{m2} \ \dots \ \mathbf{w}_{mN}] \in \mathbb{C}^{L\times N}$ is the noise matrix.

All the received signals at APs are passed to the CPU for joint activity detection. Let the collection of signals be
\begin{equation}
\scriptsize
\label{eqn:Y_CPU}
\begin{aligned}
\mathbf{Y} & = \left[ 
\begin{matrix}
\mathbf{Y}_1 \\ \mathbf{Y}_2  \\ \vdots \\ \mathbf{Y}_M
\end{matrix}
\right]  = 
\left[ 
\begin{matrix}
\mathbf{SD_aD}_{\boldsymbol{\rho}}^\frac{1}{2}\mathbf{G}_1 \\ \mathbf{SD_aD}_{\boldsymbol{\rho}}^\frac{1}{2}\mathbf{G}_2  \\ \vdots \\ \mathbf{SD_aD}_{\boldsymbol{\rho}}^\frac{1}{2}\mathbf{G}_M 
\end{matrix}
\right] + \mathbf{W}
\\
& = 
\left[  \begin{matrix}
\mathbf{S} & \mathbf{0} & \dots & \mathbf{0} \\
\mathbf{0} & \mathbf{S} & \dots & \mathbf{0} \\
\vdots & \vdots & \ddots & \vdots \\
\mathbf{0} & \mathbf{0} & \dots & \mathbf{S} 
\end{matrix}
\right]
\left[  \begin{matrix}
\mathbf{D_aD}_{\boldsymbol{\rho}}^\frac{1}{2} & \mathbf{0} & \dots & \mathbf{0} \\
\mathbf{0} & \mathbf{D_aD}_{\boldsymbol{\rho}}^\frac{1}{2} & \dots & \mathbf{0} \\
\vdots & \vdots & \ddots & \vdots \\
\mathbf{0} & \mathbf{0} & \dots & \mathbf{D_aD}_{\boldsymbol{\rho}}^\frac{1}{2} 
\end{matrix}
\right]
\left[ 
\begin{matrix}
\mathbf{G}_1 \\ \mathbf{G}_2  \\ \vdots \\ \mathbf{G}_M
\end{matrix}
\right] +  \mathbf{W}, 
\end{aligned}
\end{equation} 
where $\mathbf{W} = [\mathbf{W}_1^T \ \mathbf{W}_2^T \ \dots \ \mathbf{W}_M^T ]^T$. From  (\ref{eqn:Y_CPU}), it can be seen that the columns of $\mathbf{Y}$ are independent and each column is distributed as $\mathbf{Y}(:,i) \sim \mathcal{CN}(\mathbf{0}_{LM},\mathbf{Q})$, $\forall i=1,2,\dots,N$, where $\mathbf{Q}$ is the covariance matrix given by 
\begin{equation}
\small
\mathbf{Q} = 
\left[  \begin{matrix}
\mathbf{SD}_{\boldsymbol{\gamma}}\mathbf{D}_{\boldsymbol{\beta}_1}\mathbf{S}^H & \mathbf{0}_L & \dots & \mathbf{0}_L \\
\mathbf{0}_L & \mathbf{SD}_{\boldsymbol{\gamma}}\mathbf{D}_{\boldsymbol{\beta}_2}\mathbf{S}^H & \dots & \mathbf{0}_L \\
\vdots & \vdots & \ddots & \vdots \\
\mathbf{0}_L & \mathbf{0}_L & \dots & \mathbf{SD}_{\boldsymbol{\gamma}}\mathbf{D}_{\boldsymbol{\beta}_M}\mathbf{S}^H
\end{matrix}
\right] + \sigma^2\mathbf{I}_{LM}
\end{equation}
where $\mathbf{D}_{\boldsymbol{\beta}_m}$ is a diagonal matrix with diagonal elements corresponding to the large-scale fading coefficient from $K$ users to $m^{th}$ AP, i.e., $\mathbf{D}_{\boldsymbol{\beta}_m} = \diag(\boldsymbol{\beta}_m)$ where $\boldsymbol{\beta}_m=(\beta_{m1},\beta_{m2},\dots,\beta_{mK})$ and $\mathbf{D}_{\boldsymbol{\gamma}} = \diag(\boldsymbol{\gamma})$, where $\boldsymbol{\gamma} = (a_1\rho_1,a_2\rho_2,\dots,a_K\rho_K)$. 

By utilizing the block-diagonal structure of the covariance matrix $\mathbf{Q}$, the likelihood of $\mathbf{Y}$ given $\boldsymbol{\gamma}$ is given by
\begin{equation}
\begin{aligned}
p(\mathbf{Y}|\boldsymbol{\gamma}) & = \prod_{m=1}^{M}\prod_{n=1}^{N} \frac{1}{|\pi\mathbf{Q}_m|}\exp\left( -\mathbf{y}_{mn}^H \mathbf{Q}_m^{-1} \mathbf{y}_{mn} \right) \\
& = \prod_{m=1}^{M} \frac{1}{|\pi\mathbf{Q}_m|^N}\exp(-\trace(\mathbf{Q}_m^{-1}\mathbf{Y}_m\mathbf{Y}_m^H))
\end{aligned}
\end{equation}
where $ \mathbf{Q}_m = \mathbf{SD}_{\boldsymbol{\gamma}}  \mathbf{D}_{\boldsymbol{\beta}_m}\mathbf{S}^H + \sigma^2\mathbf{I}_L$. The maximum likelihood estimate of $\boldsymbol{\gamma}$ can be found by maximizing $p(\mathbf{Y}|\boldsymbol{\gamma})$ or equivalently minimizing $-\log(p(\mathbf{Y}|\boldsymbol{\gamma}))$ which is given by 
\begin{equation}
\label{eqn:ML_func}
\begin{aligned}
\boldsymbol{\gamma}^* = & \ \underset{\boldsymbol{\gamma}}{\arg\min} \sum_{m=1}^{M} \log|\mathbf{Q}_m| + \trace\left(\mathbf{Q}_m^{-1}\frac{\mathbf{Y}_m\mathbf{Y}_m^H}{N}\right) \\
& \text{subject to } \boldsymbol{\gamma} \geq \mathbf{0}_K
\end{aligned}
\end{equation}

To perform the activity detection, the CPU needs to solve the optimization problem in (\ref{eqn:ML_func}). Brute force approach to solve (\ref{eqn:ML_func}) requires huge complexity and the complexity increases exponentially with $M$.  In this paper, we follow the coordinate descent approach in \cite{haghighatshoar2018new} and propose an algorithm for the device activity detection that has affordable complexity, while making use of information obtained at all access points.

\section{Device Activity Detection}
\label{sec:DeviceActivityDetectionAlgorithm}
Covariance-based coordinate descent algorithm is proposed in \cite{haghighatshoar2018new} for device activity detection in co-located massive MIMO. 
In this section, we extend the coordinate descent approach for cell-free massive MIMO. We find an approximate expression for coordinate wise optimization for the ML cost function (\ref{eqn:ML_func}) and propose an algorithm for the device activity detection. 

Let $f(\boldsymbol{\gamma}) = \sum_{m=1}^{M}f^m(\boldsymbol{\gamma})$ be the cost function which needs to be optimized in (\ref{eqn:ML_func}) where $f^m(\boldsymbol{\gamma}) = \log|\mathbf{Q}_m| + \trace\left(\mathbf{Q}_m^{-1}\frac{\mathbf{Y}_m\mathbf{Y}_m^H}{N}\right)$ is the cost function  of the $m^{th}$ block in (\ref{eqn:ML_func}). Setting $\mathbf{Q}_m$ as a function of ${\boldsymbol{\gamma}}$, i.e., $\mathbf{Q}_m(\boldsymbol{\gamma}) = \mathbf{S}\mathbf{D}_{\boldsymbol{\gamma}}\mathbf{D}_{\boldsymbol{\beta}_m}\mathbf{S}^H + \sigma^2\mathbf{I}_L = \sum_{k=1}^{K}\gamma_k\beta_{mk}\mathbf{s}_k\mathbf{s}_k^H +  \sigma^2\mathbf{I}_L$, we can see $\mathbf{Q}_m$ as a sum of $K$ rank-one updates to $\sigma^2\mathbf{I}_L$. Thus we can optimize $f(\boldsymbol{\gamma})$ with respect to one argument $\gamma_k, k\in \{1,2,\dots,K\}$ in one step and we iterate several times over the whole set of variables until convergence is obtained. 
For $k\in\{1,2,\dots,K\}$, let us define $f_k^m(d) = f^m(\boldsymbol{\gamma}+d\mathbf{e}_k)$, where $\mathbf{e}_k$ is the $k^{th}$ canonical basis with a single-1 at the $k^{th}$ coordinate. 
By applying Sherman-Morrison rank-1 update identity on $\mathbf{Q}_m$ we can obtain 
\begin{equation}
\small
\label{eqn:Rank1Update}
\left(\mathbf{Q}_m + d \beta_{mk}\mathbf{s}_k\mathbf{s}_k^H \right)^{-1}  = \mathbf{Q}_m^{-1} - d\beta_{mk}\frac{\mathbf{Q}_m^{-1}\mathbf{s}_k\mathbf{s}_k^H\mathbf{Q}_m^{-1}}{1+d\beta_{mk}\mathbf{s}_k^H\mathbf{Q}_m^{-1}\mathbf{s}_k}.
\end{equation} 

Also by applying the determinant identity, we can obtain
\begin{equation}
\label{eqn:DeterminantUpdate}
|\mathbf{Q}_m + d\beta_{mk}\mathbf{s}_k\mathbf{s}_k^H| = (1+d\beta_{mk}\mathbf{s}_k^H\mathbf{Q}_m^{-1}\mathbf{s}_k)|\mathbf{Q}_m|.
\end{equation}

Now we can write the overall ML cost function in (\ref{eqn:ML_func}) for each coordinate $k$ as $f_k({d}) = \sum_{m=1}^{M}f_k^m(d)$ and is given by
\begin{equation}
\label{eqn:OverallCostFunction}
f_k({d}) = c  + \sum_{m=1}^{M}  \left( \begin{aligned}
& \log(1+d\beta_{mk}\mathbf{s}_k^H\mathbf{Q}_m^{-1}\mathbf{s}_k) \\ & \ \  -d\beta_{mk}\frac{\mathbf{s}_k^H\mathbf{Q}_m^{-1}\mathbf{Q}_{\mathbf{Y}_m}\mathbf{Q}_m^{-1}\mathbf{s}_k}{1+d\beta_{mk}\mathbf{s}_k^H\mathbf{Q}_m^{-1}\mathbf{s}_k}
\end{aligned}  \right)
\end{equation}
where $c=\sum_{m=1}^{M}\left( \log|\mathbf{Q}_m| + \trace(\mathbf{Q}_m^{-1}\mathbf{Q}_{\mathbf{Y}_m})\right)$ is a constant and $\mathbf{Q}_{\mathbf{Y}_m} = \frac{\mathbf{Y}_m\mathbf{Y}_m^H}{N}$. Finding the value of $d$ which minimizes (\ref{eqn:OverallCostFunction}) requires huge complexity and involves solving polynomials of degree $2M$. Also, as the power levels are real valued, the minimizer of (\ref{eqn:OverallCostFunction}) need not be real and can cause bad performance. Thus it calls for a low complexity design to ensure scalability of the device activity detection in cell-free massive MIMO networks. 

In the cell-free network, where the AP's and devices are distributed over large area, the large-scale fading coefficients of the device varies significantly in magnitude between different APs, unless the APs are equidistant from the device.  This variation can be up to the order of 50 dB in the presence of shadow fading. Motivated by this, at the CPU we minimize the cost function with respect to the most dominant AP for the device $k$ and the soft information about the device $k$ from this AP is propagated to other APs. 
Let $m' = \underset{m}{\argmax} \{\beta_{mk} \} $ be the access point with which the device $k$ have the dominant large-scale fading coefficient. The cost function of the device $k$ with respect to dominant AP $m'$ is given by
\begin{equation}
\label{eqn:RedefinedCostFunction}
f^{m'}_k(d) = \left( \begin{aligned}
& \log(1+d\beta_{m'k}\mathbf{s}_k^H\mathbf{Q}_{m'}^{-1}\mathbf{s}_k) \\
& \ \ \ -d\beta_{m'k}\frac{\mathbf{s}_k^H\mathbf{Q}_{m'}^{-1}\mathbf{Q}_{\mathbf{Y}_{m'}}\mathbf{Q}_{m'}^{-1}\mathbf{s}_k}{1+d\beta_{m'k}\mathbf{s}_k^H\mathbf{Q}_{m'}^{-1}\mathbf{s}_k}
\end{aligned}
\right).
\end{equation}
Taking the derivative of (\ref{eqn:RedefinedCostFunction}) and setting it to zero, we obtain 
\begin{equation}
d^* = \frac{\mathbf{s}^H_k\mathbf{Q}^{-1}_{m'}\mathbf{Q_{Y}}_{m'}\mathbf{Q}^{-1}_{m'}\mathbf{s}_k - \mathbf{s}^H_k\mathbf{Q}^{-1}_{m'}\mathbf{s}_k }{\beta_{m'k}(\mathbf{s}^H_k\mathbf{Q}^{-1}_{m'}\mathbf{s}_k)^2}.
\end{equation}

To preserve the positivity of $\boldsymbol{\gamma}$ in (\ref{eqn:ML_func}), the optimal update step $d$ is given by $d=\max\{d^*,-\gamma_k\}$ and the coordinate is updated as $\gamma_k=\gamma_k+d$. This update step $d$ is propagated to all the sub covariance matrices $\mathbf{Q}_m$'s and are updated using (\ref{eqn:Rank1Update}). This optimization will be done over whole set of random permutation of variables from set $\{1,2,\dots,K\}$ and we iterate the entire procedure until we obtain convergence. The proposed algorithm is illustrated in Algorithm 1. The complexity of the proposed algorithm is $\mathcal{O}(TKML^2)$, where $T$ is the number of iterations.

\floatstyle{spaceruled}
\restylefloat{algorithm}
\begin{algorithm}[!t]
	\caption{\strut Coordinate Descend Algorithm for estimating $\boldsymbol{\gamma}$}
	\begin{algorithmic}[1]
		\renewcommand{\algorithmicrequire}{\textbf{Input:}}
		\renewcommand{\algorithmicensure}{\textbf{Initialize:}}
		\REQUIRE Observations $\mathbf{Y}_m, \forall m=1,2,\dots M$, $\beta_{mk}, \forall m=1,2,\dots M , k=1,2,\dots K$
		\ENSURE  $\mathbf{Q}^{-1}_m=\sigma^{-2}\mathbf{I}_L ,  \forall m=1,2,\dots M$, $\hat{\boldsymbol{\gamma}}=\mathbf{0}_K$
		%		\\ \textit{Initialisation} :
		\STATE Compute $\mathbf{Q_{Y}}_m = \frac{1}{N}\mathbf{Y}_m\mathbf{Y}^H_m,  \forall m=1,2,\dots M$
		%		\\ \textit{LOOP Process}
		\FOR {$i = 1,2,\dots,T$ }
		\STATE Select an index set $\mathcal{K}$ from the random permutation of set $\{1,2,\dots,K\}$
		\FOR {$k\in\mathcal{K}$}
		\STATE Find the strongest link or AP for device $k$ , i.e., \\ $m' = \underset{m}{\argmax} \{\beta_{mk} \} $
		\STATE $\delta = \max \left\{ \frac{\mathbf{s}^H_k\mathbf{Q}^{-1}_{m'}\mathbf{Q_{Y}}_{m'}\mathbf{Q}^{-1}_{m'}\mathbf{s}_k - \mathbf{s}^H_k\mathbf{Q}^{-1}_{m'}\mathbf{s}_k }{\beta_{m'k}(\mathbf{s}^H_k\mathbf{Q}^{-1}_{m'}\mathbf{s}_k)^2},-\hat{\gamma}_k \right\} $
		\STATE$\hat{\gamma}_k \leftarrow \hat{\gamma}_k + \delta$
		\FOR {$m=1,2,\dots,M$}
		\STATE $\mathbf{Q}^{-1}_m \leftarrow \mathbf{Q}^{-1}_m - \delta\frac{\beta_{mk}\mathbf{Q}^{-1}_m \mathbf{s}_k \mathbf{s}^H_k \mathbf{Q}^{-1}_m}{1+\delta\beta_{mk}\mathbf{s}^H_k\mathbf{Q}^{-1}_m\mathbf{s}_k}$
		\ENDFOR
		\ENDFOR
		\ENDFOR
		\RETURN $\hat{\boldsymbol{\gamma}}$
	\end{algorithmic}
\end{algorithm}

To perform activity detection, the output from Algorithm 1 is compared against a threshold $\gamma_k^{th}$ for each device $k$ and is given by 
\begin{equation}
\label{eqn:ActivityDetection}
\hat{a}_k = \left\{ \begin{aligned}
1, & \text{ if } \hat{\gamma}_k \geq \gamma_k^{th} \\
0, & \text{ otherwise}.
\end{aligned} \right. 
\end{equation}
The threshold $\gamma_k^{th} = \nu\frac{\sigma^2}{\beta_{m'k}}$ where $\nu>0$ is chosen to have a desired probability of miss detection and probability of false alarm performance. 

\section{Simulation Results}
\label{sec:SimulationResults}
In this section, we characterize the massive connectivity in distributed MIMO architectures and plot the performance of massive activity detection in cell-free massive MIMO with our proposed algorithm.
\subsection{Performance Metrics}
%The network access probability is defined as the average ratio of number of active users granted access to the number of active users and is given by
%$$P_a = \mathbb{E}\left\{ \frac{|\mathcal{A}_G|}{|\mathcal{A}|} \right\}$$
%where $\mathcal{A}_G = \{k : a_k=1,\text{SNR}_k\geq \text{SNR}_{th}\}$ is the set of active users which met the SNR or power constraints. $\text{SNR}_k$ and $\text{SNR}_{th}$ are SNR of user $k$ and the SNR threshold to be maintained at the receiver respectively.

We consider the receiver operating characteristic (ROC) as the performance measure for activity detection. Let $\hat{\mathcal{A}}=\{k \ | \ \hat{a}_k=1, \forall k\in [1,K] \ \}$ be the estimate of the set of active devices. The probability of miss detection is defined as the average of the ratio of non-detected devices and the number of active devices and the probability of false alarm is defined as the average of inactive devices declared active over inactive devices and are given by
\begin{equation}
P_{md} = 1-\mathbb{E}\left\{\frac{|\mathcal{A}\cap\hat{\mathcal{A}}|}{|\mathcal{A}|}\right\} , \ \ P_{fa} = \mathbb{E}\left\{\frac{|\hat{\mathcal{A}} \setminus \mathcal{A} |}{K-|\mathcal{A}|}\right\}.
\end{equation}

\begin{figure}[t]
	\centering
	\includegraphics[scale=0.57]{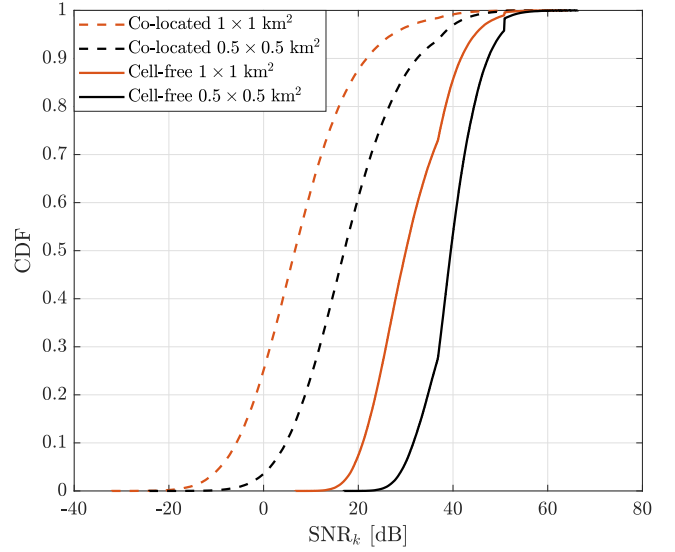}
	\caption{Active device SNR}
	\label{fig:ActiveDeviceSNR}
\end{figure}

\subsection{Simulation Model}
We consider a square area wrapped around the edges to imitate a network with infinite area and to avoid boundary effects where the $M$ AP's and $K$ users are uniformly distributed at random. For co-located case, we consider the AP is at the center of network. We consider such a cell area with $K=400$ devices, the activation probability $\epsilon=0.1$ and the signature sequence length $L=40$. The following three slope propagation model used in \cite{ngo2017cell} is considered for large-scale fading coefficient $\beta_{mk}$:
\begin{equation}
\small 
\label{eqn:PathLoss}
\beta_{mk}[\text{dB}] = \left\{ 
\begin{aligned}
&-81.2 && d_{mk} < 10  \\ 
&-61.2 -20\log_{10}(d_{mk}) && 10\leq d_{mk} < 50  \\
&-35.7-35\log_{10}(d_{mk}) + F_{mk} && d_{mk} \geq 50 
\end{aligned}  \right.
\end{equation} 
where $F_{mk}\sim\mathcal{N}(0,8^2)$ is the shadow fading component and $d_{mk}$ is the distance between $k^{th}$ user and $m^{th}$ AP in meters. The maximum transmit power for a device is 200 mW, the bandwidth is 1 MHz and noise power $\sigma^2=-109 \ \text{dBm}$. 

\subsection{Results}
First, we compare the received SNR at the base station antenna for an active device $k$ transmitting at a power of 200 mW for co-located and cell-free massive MIMO networks with different cell sizes. Fig. \ref{fig:ActiveDeviceSNR} shows that there is a significant gap in the received SNR for co-located and cell-free MIMO and hence the outage probability is less for cell-free MIMO scenario.

\begin{figure}[t]
	\centering
	\includegraphics[scale=0.61]{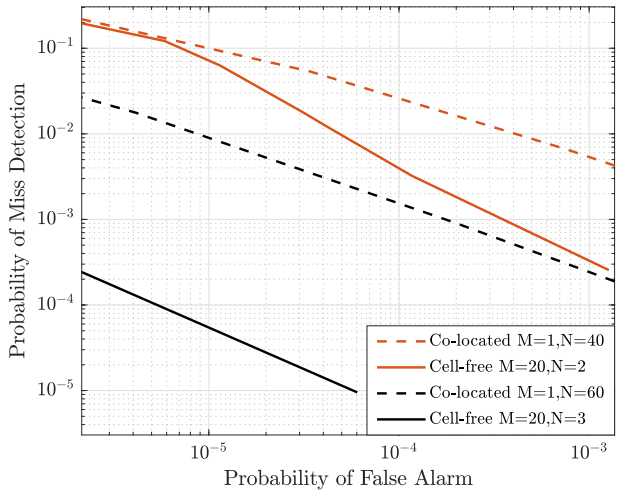}
	\caption{Performance in $1\times1 \ \text{km}^2$ cell area. $K=400$ users, activation probability $\epsilon=0.1$, sequence length $L=40$}
	\label{fig:Performance_1sqkm}
\end{figure}

\begin{figure}[t]
	\centering
	\includegraphics[scale=0.6]{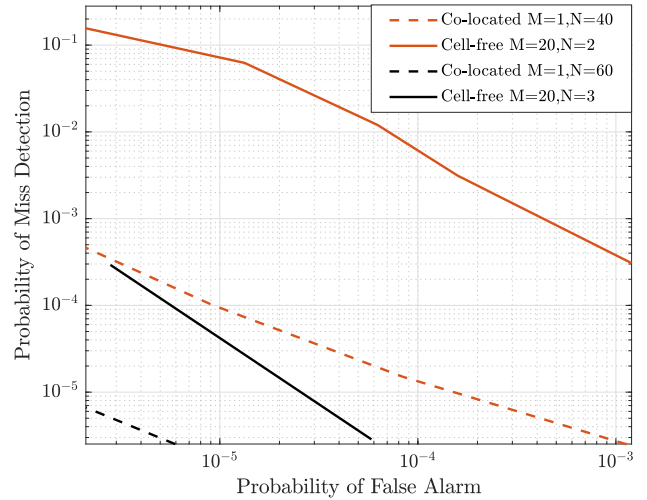}
	\caption{Same as Figure \ref{fig:Performance_1sqkm} but for $500\times500 \ \text{m}^2$ cell area.}
	\label{fig:Performance_500sqm}
\end{figure}

Next we look at the performance of proposed algorithm for activity detection in cell-free massive MIMO network. For simulations, we consider a SNR target at the base station such that 95\% of the active devices will be able to achieve the desired SNR and hence access the network. The ROC curve is plotted for different choices of threshold. We have considered $T=10$ as the number of iterations in Algorithm 1. The performance is given in fig. \ref{fig:Performance_1sqkm} and fig. \ref{fig:Performance_500sqm} for $1\times1 \  \text{km}^2$ and $0.5\times0.5 \ \text{km}^2$ cell area sizes respectively. For low-power applications like mMTC, co-located MIMO is highly sensitive to receive SNR and performance degrades significantly with increase in cell area. It can be seen that device activity detection performance is robust against shadow fading effects in cell-free massive MIMO networks compared to co-located MIMO networks and hence cell-free network can provide a broad coverage in mMTC applications. The performance in cell-free massive MIMO network significantly improves with increase in number of antennas per AP. 

\section{Conclusion}
\label{sec:Conclusion}
In this paper, we studied about the grant-free random access scenario in cell-free massive MIMO networks. The paper formulates activity detection problem in cell-free massive MIMO network and provides an approximate solution to the estimator. An algorithm based on coordinate descent is provided for device activity detection with affordable complexity. We show that for low-powered applications like mMTC, co-located massive MIMO is highly sensitive to receive SNR while cell-free massive MIMO is robust against the shadow fading effects and hence can provide better coverage. A direction for future work is to find the soft information of each device from cluster of dominant APs instead of the most dominant AP. 

\bibliographystyle{IEEEtran}
\bibliography{IEEEabrv,references}

\end{document}